\documentclass[aip,jcp,preprint,showkeys,superscriptaddress,longbibliography,noeprint]{revtex4-1}
\usepackage[utf8]{inputenc}
\usepackage[english]{babel}
\usepackage[utf8]{inputenc}
\usepackage{systeme}
\usepackage[fleqn]{amsmath}
\usepackage{amssymb}
\usepackage[separate-uncertainty=true, multi-part-units=single]{siunitx}
\usepackage[inline,shortlabels]{enumitem}

\newcommand{\red}[1]{{\color{black} #1}}

\newcommand{\green}[1]{{\color{black} #1}}

\newcommand{\VEC}[1]{\boldsymbol{#1}}

\usepackage{graphicx,grffile,subfigure}
\graphicspath{{figures/}}
\newlength\figwidth
\usepackage{url,hyperref}
\usepackage[table,usenames,dvipsnames]{xcolor}
\hypersetup{colorlinks=true, linkcolor=BrickRed,
  urlcolor=blue!50!black, citecolor=blue!50!black}
\usepackage[capitalize]{cleveref}
\crefrangelabelformat{equation}{(#3#1#4)$-$(#5#2#6)}

\makeatletter
\def\paragraph{%
  \@startsection
    {paragraph}{4}{\parindent}{\z@}{-1.5em}%
    {\normalfont\normalsize\itshape}%
}%
\makeatother

\newcommand{\eq}[1]{(\ref{#1})}
\newcommand{\reals}{\mathbb{R}}

\newcommand{\VXi}{\VEC{\Xi}}
\newcommand{\VX}{\VEC{X}}
\newcommand{\VY}{\VEC{Y}}
\newcommand{\vF}{\vec{F}}
\newcommand{\vn}{\vec{n}}
\newcommand{\vp}{\vec{p}}
\newcommand{\vv}{\vec{v}}
\newcommand{\Vp}{\VEC{p}}
\newcommand{\vq}{\vec{q}}
\newcommand{\Vq}{\VEC{q}}

\newcommand{\Vtotal}{V_{\text{tot}}}
\newcommand{\Fmean}{\vec{F}_{\text{av}}}

\newcommand{\fouter}{f^{\circ}}

\newcommand{\dss}{\displaystyle}

\begin{document}
\title{Liouville-type equations for the $n$-particle distribution functions of an open system}

\newcommand\FUBaffiliation{\affiliation{Freie Universität Berlin, Institute of Mathematics, Arnimallee 6, 14195 Berlin, Germany}}
\author{Luigi Delle Site}
\email{luigi.dellesite@fu-berlin.de}
\FUBaffiliation
\author{Rupert Klein}
\email{rupert.klein@math.fu-berlin.de}
\FUBaffiliation

\begin{abstract}
  \today\\
  {In this work we derive a mathematical model for an open system that exchanges particles and momentum with a reservoir from their joint Hamiltonian dynamics. The complexity of this many-particle problem is addressed by introducing a countable set of $n$-particle phase space distribution functions just for the open subsystem, while accounting for the reservoir only in terms of statistical expectations. From the Liouville equation for the full system we derive a set of coupled Liouville-type equations for the $n$-particle distributions by marginalization with respect to  reservoir states. The resulting equation hierarchy describes the external momentum forcing of the open system by the reservoir across its boundaries, and it covers the effects of particle exchanges, which induce probability transfers between the $n$- and $(n+1)$-particle distributions. Similarities and differences with the Bergmann-Lebowitz model of open systems {(P.G.\ Bergmann, J.L.\ Lebowitz, Phys.~Rev., 99:578--587 (1955))} are discussed in the context of the implementation of these guiding principles in a computational scheme for molecular simulations.}
\end{abstract}
\maketitle
\section{Introduction}
Open systems of microscopic particles that exchange energy and matter with a large environment are ubiquitous in the physical world. {Strictly speaking, closed systems that exchange neither particles nor momentum or energy with their environment don't even exist.} Examples {of open systems} span from naturally-occurring systems to artificially engineered systems of modern and future technology\cite{quianrev,bordin,abramo,politzer,herzog,advtsel}. The description of the complexity of such systems {often} requires computer simulations for the prediction of {their} physical properties. 
{Yet, the computational study of open systems comes with a higher degree of complexity than that of closed ones: Energy/momentum and particle exchanges with the environment render obsolete the established computational protocols, in particular those of Molecular Dynamics (MD), which rely on just these conservation laws by design \cite{Frenkelbook,tuckbook}.} 

{Meanwhile, the latest generation of MD algorithms} based on the concept of adaptive molecular resolution provide a generic root model for computational approaches to open systems \cite{adress2,ensing,truhlar,wagoner,csany,physrep,cpcel,softmatt}: Different regions of space are treated at different molecular resolution while allowing molecules/particles to freely move in space and change resolution accordingly. As a consequence, one has a prototype scheme for simulating a high resolution region coupled to a simplified environment (reservoir) with the exchange of energy and matter.

Applications of such schemes to real world systems led to accurate numerical results for a large class of physio- and bio-chemical systems \cite{ensingpol,mat1,mat2,krem1,krek1,krem2,shad1,shad2,shad3,mat3,pandemonium}. In this perspective a mathematical formalization of the model is {highly desirable as it helps to avoid situations in which largely} empirical definitions of calculated quantities lead to artificial results {or} to their misleading interpretation. For example, a key question posed in {recent years regards} the proper definition of time correlation functions in the high resolution region for {those} degrees of freedom that are not present in the simplified reservoir \cite{njp,jcppi,preliou,cpcpi,pccppi}. The definition of a {proper time evolution equation for the phase space probability densities, i.e., the analogue of a Liouvillian, is mandatory in this case. In other words, a mathematical model that formalizes the algorithms and makes their implied probability density evolution explicit became necessary.} 

In this perspective, {the Adaptive Resolution Simulation scheme (AdResS) \cite{adress1} has been} mapped onto a formal model of open systems known as the Bergmann-Lebowitz model \cite{leb1,leb2}. {This} model, whose details are reported later on in this paper, defines a Liouville-type equation for the probability distribution of the open system via a stochastic coupling with the reservoir based on the idea of impulsive {system-reservoir interactions}. The mapping between AdResS and the Bergmann-Lebowitz model is unfortunately only qualitative \cite{njp}, but it {provides an attractive reference for the structure of} the algorithm and {led to favorable comparisons of time correlations between AdResS and fully atomistic simulations  \cite{softmatt}}. The key characteristic of AdResS and of similar {schemes is an exchange of particles and momentum/energy with {some reservoir} according to classical Newtonian dynamics. In contrast,} the Bergmann-Lebowitz model features a stochastic coupling {to describe this exchange}.  {The aim of the present work is to derive a mathematical model that is similar in spirit to the Bergmann-Lebowitz model for open systems but based on the first principles of Newtonian mechanics, and that therefore accounts for interactions with the reservoir as they occur in the computational model}. 

{We propose to describe the evolution of the probability distribution of the open system by introducing a total of $N$ $n$-particle phase space distribution functions $f_n$, where $N$ is the number of particles in the entire System and $n \in \{0, ..., N\}$ denotes one of the possible numbers of particles that reside in the considered open subsystem at any instance of time. A hierarchy of Liouville-type equations, one for each of the $f_n$,} is derived directly from the full Liouville equation for the {probability distribution $F_N$ of the entire $N$-particle system as a whole. The derivation proceeds by integrating out, i.e., marginalizing w.r.t., the degrees of freedom of the $N-n$ particles that reside outside the open system subregion}.  

{The result is a coupled system of Liouville-type equations for the entire hierarchy of distributions $(f_n)_{n=0}^N$. The coupling arises naturally in the form of external forcing across the boundaries of the open system, and it is due to the exchange of particles and momentum/energy with the outside reservoir. A solution of this coupled system describes the time evolution of all $n$-particle probability distributions of the open system simultaneously. The probabilistic exchange of particles with the reservoir is accounted for by the probabilities of transitions from $n$- to $n+1$-particle states across the hierarchy of equations. Thus, at each time $t$ the set of distributions $(f_n)_{n=0}^N$ characterizes the state of the open system in the sense of statistical mechanics}. 

{If one is not interested in the detailed particle-based description of the reservoir, then the key theoretical step is the realization that in absence of detailed information regarding the time evolution of the reservoir state, probability distributions for the state of the open system can only be described by statistical averages over the reservoir's microscopic degrees of freedom}. Adopting the simplest possible approach, the derivation in this paper considers the \emph{expectations} of the $n$-particle open system distributions \emph{with respect to the statistics of the reservoir}.

The key point {in connecting this model with the AdResS computational approach} is the introduction of a boundary/surface layer {around the open system whose particles may enter the system or exit fully into the reservoir, and which are close enough to the open system boundary that their pair potential momentum exchange with particles inside the system is non-negligible during a time step. This boundary layer, which is the central signature of the AdResS scheme, is responsible for modelling the statistics of the outside reservoir which enters the theory only through its single- and two-particle distributions}.

We follow a procedure similar to the so-called Bogoliubov-Born-Green-Kirkwood-Yvon hierarchy scheme (BBGKY model) \cite{bog3}, where the Liouville equation in a system of $N$ particles is written for a subset $s$ of particles whose probability distribution function is the $N-s$-marginalized probability distribution function of the total system. However, our equations are substantially different in form and meaning from those of the BBGKY model, due to the {partitioning of position space in our model into the domain of the open system and its complement.} 

The paper is organized as follows: First we define the terms of the problem, that is we discuss the Liouville equation for a hierarchy of  $N$-particle distribution functions specifically aimed at the description of an open system. Next we show how this {hierarchy} is constructed and present the explicit expression of the resulting Liouville{-type equations}. A verification for its physical consistency is provided by discussing the situation of equilibrium and the resulting Grand Canonical distribution. To stress the difference between the proposed approach and fully stochastic system-reservoir coupling models, we discuss {similarities and differences} with the Bergmann-Lebowitz model \cite{leb1,leb2}. Finally, the model is put in relation with computational approaches aiming at simulating open systems, in particular we will discuss the correspondence between our model and the principles of the AdResS algorithm.
\section{Liouville-type equations for the $N$-particle distribution functions of an open system}
\label{sec:OpenSystemModel}
\subsection{Liouville-type equations}
In general, a Liouville equation
is the mathematical formalization of the concept that the statistical weight of an individual member of an ensemble of realizations of a dynamical system does not change in time. 
For a closed system, this is equivalent to stating that the probability density of initial 
states of the system is advected along trajectories of the dynamics, and, for an $N$-particle
system, it is expressed by the phase space density transport equation:
\begin{equation}\label{eq:LiouvilleTypeEquationClosedSystem}
\frac{\partial f(t,{\Vq},{\Vp})}{\partial t} 
+ \sum_{i=1}^{N} \left(v_i \cdot \nabla_{q_i}  
+ F_i \cdot \nabla_{p_i} \right) f(t,{\Vq},{\Vp})
= 0\,.
\end{equation}
with $f(t,{\Vq},{\Vp})$ the distribution function in the $6N$-dimensional $({\Vq},{\Vp})$ phase space system, $v_{i}$ the velocity and $F_{i}$ the force {acting upon} the $i$-th particle, and analogously {$(\vq_{i},\vp_{i})$ the $6$-dimensional position and momentum phase space} point of the  of the $i$-th particle. In this work we shall go beyond the standard application of Eq.~\ref{eq:LiouvilleTypeEquationClosedSystem} and explore {its consequences for the statistics of a finite size open subsystem which, in the course of time, may host in principle any number of particles from $0$ to $N$ as part of the dynamics. We will therefore have to address the time evolution of an entire hierarchy of $n$-particle distributions $f_n(t,\VX^n)$ \ $(n = 0, ..., N)$.}

{Notice that this hierarchy will differ from the well-known BBGKY hierarchy for $n$-particle marginal distributions of the original system in that we restrict to $n$ particle distributions conditioned on their {and only their} residence in the open system subdomain $\Omega$.}



\subsection{General set-up for a hierarchy of phase space density functions}
As underlined before, the interest of this work is in the mathematical modelling of an open system of particles subject to Hamiltonian dynamics internally and subject to external forcing from its boundaries. The Hamiltonian for a situation that has $n$ particles in the system's spatial domain $\Omega$ reads: 
\begin{equation}\label{eq:Hamiltonian}
H_n 
= \sum_{i=1}^{n} \frac{\vp_i^2}{2M} 
+ \sum_{i=1}^{n} \sum_{j\not=i}^{n} \frac{1}{2} V(\vq_j-\vq_i) \qquad (\vq_i, \vq_j \in \Omega)
\end{equation}
where $V(\vq)$ is a two-particle potential, $X_i \equiv (\vq_i,\vp_i) \in S = \Omega \times \reals^3$ are the $i$th particle's physical and momentum space coordinates, respectively, and $M$ is the mass of an individual particle.

By saying that the system is ``open'' we mean that it can exchange particles with a{, typically  large,} surrounding reservoir. Furthermore, we aim to account for the possibility that particles which reside inside the system's boundary can interact with reservoir particles outside via inter-particle forces. These interactions are responsible for an exchange of kinetic energy{/momentum} between the system and its surroundings.

By definition of an open system, details of the ``state of the world'' at any time are known only for the particles inside the system. Properties of the ``rest of the world'' have to be assumed or prescribed by the modeller, and they depend on which kind of embedding she or he is thinking of. In the present work we assume that the outside world is described by suitable statistical information. That being the case, the open system description to be developed can only be statistical with respect to the outside world, the simplest case being a description in terms of expectations. 

In this sense, the present suggestion of a ``Liouville-type'' system of equations for open systems differs substantially from the classical Liouville equation for a closed particle system with deterministic dynamics: the evolution of phase space densities is, in this latter case, fully deterministic, while here we have in principle even a statistical ensemble of hierarchies of open system phase space densities. Restricting to expectations relative to the outside world, we reduce this complexity to the task of handling the evolution of just a finite, or in the worst case countable, set of $n$-particle phase space distribution functions. {This sequence of probability distributions is similar to the BBGKY hierarchy of $n$-particle probability densities with the key distinction that we consider $n$-particle densities conditional upon $n$ and only $n$ particles residing inside the open system's domain.}  

A mathematical characterization of the system is given through an extended Liouville equation  for the probability density:
%
\begin{equation}\label{eq:DensityHierarchy}
\begin{array}{rcrcl}
f_n
  & : 
    & \reals^+ \times S^n  
      & \to 
        & \reals
          \\
  &
    & (t, \VX^n) 
      & \mapsto 
        & f_n(t, \VX^n) \qquad\text{for}\qquad (n = 0, ..., N)\,,
\end{array}
\end{equation}
with
\begin{equation}
\VX^n = [X_1, ..., X_n] \qquad\text{and}\qquad X_i = (\vq_i,\vp_i) \in S = \Omega \times \reals^3\,,
\end{equation}
of finding, at time $t$, any $n$ particles in the system with locations $(X_i)_{i=1}^n$ in the particles' phase space. {For a ``universe'' with a total of $N$ particles in a domain $U$ that contains $\Omega$ and that has an $N$-particle density $F_N(t, \VX^N)$ for \emph{identifiable particles}, we let} \\
%
\begin{equation}\label{eq:HierarchyDefinition}
f_n(t, \VX^n) 
= {{N}\choose{n}}
  \int\limits_{(S^c)^{N-n}}  
  F_N(t, \VX^n, \VXi_n^{N})\ d\VXi_n^{N} \qquad
  \text{where}\qquad S^c = \Omega_c \times \reals^3
\end{equation}
with $\Omega_c = U \backslash \Omega$ {and 
\begin{equation}
\VXi_n^{N}  \equiv [\Xi_{n+1},.....{\green{\Xi_{N}}}] 
\qquad\text{where}\qquad
\Xi_{i}=(\vq_i,\vp_i) \in S^c\,.
\end{equation}
} The combinatorial prefactor in \eq{eq:HierarchyDefinition} results from the fact that the (classical) particles are indistinguishable, so that the probability density $F_N$ is symmetric w.r.t\ any permutation of the particles. For future reference, we let 
\begin{equation}\label{eq:JustMarginalizedF}
f^*_n(t, \VX^n) = {{N}\choose{n}}^{-1} f_n(t, \VX^n) 
\end{equation}
denote the full $N$-particle distribution marginalized with respect to its last $N-n$ coordinates, and without account of the indistinghuishability of the particles.

{The distributions in \eq{eq:DensityHierarchy} form a hierarchy of coupled $n$-particle density functions since, for any realization of the system, the total number of particles inside of its domain is {generally} time dependent. As a consequence, we need distributions for any particle number $n \in \{0,1,...,N\}$ for a complete statistical description of the system.}
{The normalization condition for $F_N$, i.e., $\int_{S^N} F_N\, d\VX^N = 1$, implies the normalization condition for the $n$-particle hierarchy
\begin{equation}\label{eq:NormalizationCondition}
\sum\limits_{n = 0}^{N} \
\int\limits_{\Omega^n} \int\limits_{(\reals^{3})^{n}} f_n(t,(\Vq,\Vp))\ d\Vp\, d\Vq = 1\,.
\end{equation}
\green{The normalization condition of \eq{eq:NormalizationCondition} is routinely used in popular textbooks of statistical mechanics when partitioning a large system in (open) small subsystems, see e.g. Refs.\citenum{huangbook,tuckbook}. In section \ref{gcverification} it will be explicitly verified that \eq{eq:NormalizationCondition} is consistent with the derivation of the grand canonical distribution function.
In physics the sum is usually extended to $\infty$; such an approach is based on qualitative arguments (thermodynamic limit), and it will be used here in section \ref{gcverification}. In mathematics the extension of the sum to $\infty$ is not straightforward and it represents an interesting problem. However given the current focus, this discussion goes beyond the scope of this paper.}


\section{Derivation of the Liouville-type equation hierarchy}
Here we consider an open subsystem occupying the spacial domain $\Omega$ embedded in a much larger ``Universe'' $U \subset \reals^3$ of $N$ particles. The {starting point of the derivation is the Liouville equation for the $N$-particle distribution function. This is a standard Liouville equation for a closed system of $N$ particles and probability distribution $F_N(t,\VX^N)$, that is, the conservation law for the probability of the universe. In mathematical terms this means a definition of $F_N(t,\VX^N)$ as:
\begin{equation}
\begin{array}{rrcl}
F_N :
  &  \dss \reals^+ \times (U \times \reals^3)^N 
    & \to 
      & \dss \reals 
        \\
  &  \dss (t, \VX^N) 
    & \mapsto 
      & \dss F_N(t,\VX^N)
\end{array}\,,
\end{equation}
and the corresponding equation of conservation:

\begin{equation}\label{eq:NParticleLiouville}
\frac{d F_N}{dt} 
=0
\end{equation}
and as a consequence, in the form known as Liouville equation:
\begin{equation}
  \label{eq:liouvilleNPspecific}
\frac{\partial F_N}{\partial t}=\green{-}\sum\limits_{i=1}^N 
\left[
\nabla_{\vec{q}_i} \cdot \left({\red{\vec{v}_i}} F_N\right)
+
\nabla_{\vec{p}_i} \cdot \left(-\nabla_{q_i}\Vtotal(\Vq^N) F_N\right)
\right]
\,,
\end{equation}
where 
\begin{equation}\label{eq:TotalPotential}
\Vtotal(\Vq^N) = \sum_{i<j}^{N}V(\vq_i-\vq_j)
\end{equation}}
is the total potential energy of the system as a function of the positions of all of its $N$ particles. True to the definition of the open system $n$-particle distributions in \eq{eq:HierarchyDefinition}, we marginalize this general evolution equation w.r.t.\ $N-n$ particles residing outside of the open system subdomain $\Omega$. 



\subsection{Marginalizing w.r.t.\ outside particle momentum spaces}

{Here we consider all possible configurations conditioned upon $n$ and only $n$ particles residing within the open subsystem and integrate the Liouville equation \eq{eq:liouvilleNPspecific} over the $N-n$ degrees of freedom of the remaining particles outside the open subsystem. After this operation and by the definition of $f_n$ in \eq{eq:HierarchyDefinition}, the left hand side yields the time derivative $\frac{\partial f_n}{\partial t}$ of the $n$ particle distribution. The terms on the right of the equation will be shown below to yield the expected Liouville-type transport terms for $f_{n}$ itself as well as coupling terms between $f_n$ and other functions $f_k$ for $k \not= n$.} 

Let $\Xi_k = (\vec{q}_k, \vec{p}_k)$ for some $k$ such that ${\green{n+1}} < k \leq N$ and consider the marginalization of the term $\nabla_{\vec{p}_k}\cdot\left(\vec{F}_k\, F_N\right)$ with respect to $\vec{p}_k$. Integrating the term over any ball $B_r(0)$ of momentum space, we find
\begin{equation}\label{eq:MomentumFluxDivTermOuterParticle}
\int\limits_{B_r(0)} \nabla_{\vec{p}_k}\cdot\left(-\nabla_{q_k}\Vtotal(\Vq^N)\, F_N\right) d^3 p_k 
=
- \int\limits_{\partial B_r(0)} \vec{n}\cdot\left(\nabla_{q_k}\Vtotal(\Vq^N)\right) F_N\, d\sigma_{p_k} 
\to 0
\qquad
(r \to \infty)
\end{equation}
In the limit of the radius of the ball tending to infinity the term will generally vanish because the probability $F_N$ will decay sufficiently rapidly for large momenta. For the equilibrium distribution this is obvious, since in this case $F_N$ factorizes into the momentum and position terms, and the momentum distribution is proportional to $\exp(-(\vec{p}_k/M)^2/2kT)$. All the terms in \eq{eq:MomentumFluxDivTermOuterParticle} with ${\green{n+1}} \leq k \leq N$ follow this reasoning, so that they all vanish identically. 

In contrast, for $1 \leq i \leq n$, we have to carefully account for the dependencies of the total potential energy on the open system and universe coordinates. To this end, we assume here that the total potential is the sum of pair interactions as {described} in \eq{eq:TotalPotential}. Under this assumption, let us analyze the behavior {of} one of the terms under this double-sum when the momentum flux divergence in the Liouville equation is marginalized. If both, $i,j \leq n$, then 
\begin{equation}\label{eq:MomentumFluxDivTermTwoInnerParticles}
\int\limits_{(S^{c})^{N-n}}  
\nabla_{\vec{p}_i}\cdot\Bigl(\nabla_{q_i}V(\vq_i-\vq_j)\, F_N(t, \VX^n, \VXi_n^{N})\Bigr) d\Xi_n^N 
=
\nabla_{\vec{p}_i}\cdot\Bigl(\nabla_{q_i}V(\vq_i-\vq_j)\, f^*_n(t, \VX^n)\Bigr) \,,
\end{equation}
provided the probability density is sufficiently smooth to allow for an exchange of integration and differentiation w.r.t.\ complementary coordinates. Thus, taking into account the particle exchange symmetry again (see \eq{eq:JustMarginalizedF}), we obtain the standard momentum flux divergence contribution that we would expect in an $n$-particle Liouville equation. 

If, however, in one of the terms in \eq{eq:TotalPotential} only $\vq_i\in \Omega$ but $\vq_j \in \Omega_c$, then the marginalization of the corresponding term in the Liouville equation reads
\begin{equation}\label{eq:MomentumFluxDivTermOneInnerOneOuterParticle}
\begin{array}{rl}
\dss \int\limits_{S^c}\int\limits_{(S^{c})^{N-n-1}}  
  & \hspace{-3mm} \dss \nabla_{\vec{p}_i}\cdot
    \Bigl(-\nabla_{q_i}V(\vq_i-\vq_j)\, F_N(t, \VX^{n-1,}, X_i, (\vq_j,\vp_j), \VXi_{n+1}^{N})\Bigr) {\green{d\VXi_{n+1}^{N}}}
    \ dp_j dq_j \,.
\end{array}
\end{equation}
Due to the dependence of the force on $\vq_j$ and the fact that we are marginalizing over that variable, this term cannot be straightforwardly expressed in terms of the hierarchy of distributions $(f_n)_{n=0}^{N}$. An explicit step of modelling is required for closure. 

As this paper is to describe a general framework but is not focused on some specific system, let us suggest here one plausible closure model and leave the discussion of more sophisticated options for future work:  Suppose that the pair interaction $V(r)$ is relatively short-range in comparison with $\text{diam}(\Omega)$, so that the pair interactions are felt only close to the open system's boundary. Suppose further that the probability density of finding $n$ particles in states $\left(\VX^{n-1}, X_i\right) \in S^n$ and one other outer particle in $X_j$ is given by $f_n\left(\VX^{n-1}, X_i\right) f^{\circ}_2(X_j | X_i)$, where $f^{\circ}_2(X_{\text{out}} | X_{\text{in}})$ is a known or modelled conditional distribution for joint appearances of an outer particle given the state of an inner one. The key modelling assumption here is that an inner particle $i$ whose position $q_i$ is sufficiently close to the boundary $\partial \Omega$ for it to feel the pair interaction from the outer particles, this conditional distribution is approximately independent of the states $\VX^{n-1}$ of the other $n-1$ particles within the open system. 

Given this modelling assumption, \eq{eq:MomentumFluxDivTermOneInnerOneOuterParticle} results in a mean-field expression for the momentum flux divergence term, 
\begin{equation}\label{eq:MomentumFluxDivTermOneInnerOneOuterParticleII}
\begin{array}{rl}
\dss \int\limits_{S^c}\int\limits_{(S^{c})^{N-n-1}}  
  & \hspace{-3mm} \dss \nabla_{\vec{p}_i}\cdot
    \Bigl(-\nabla_{q_i}V(\vq_i-\vq_j)\, F_N(t, \VX^{n-1}, X_i, (\vq_j,\vp_j), \VXi_{n+1}^{N})\Bigr) 
    d\VXi_{n+1}^{N}
    \ dp_j dq_j \,.
    \\
    & \dss = \nabla_{\vec{p}_i}\cdot\Bigl(\Fmean(\vq_i) f^*_n(t, \VX^{n-1}, X_{i})\Bigr) \,,
\end{array}
\end{equation}
where
\begin{equation}
\Fmean(\vq_i) 
= 
- \int\limits_{S^c} 
  \nabla_{\vq_i}V(\vq_i- \vq_j) f^\circ_2(X_j | X_i) \,d X_j
\end{equation}
is the mean field force exerted by the outer particles onto the $i$th inner particle. The derived expression is invariant w.r.t.\ particle permutations which accounts for a multiplicity of $N!$. At the same time, we keep $f_n(t,\VX^n)$ as a symmetric function w.r.t.\ particle permutations inside the open system, so that the statistical weight of a particular configuration is only $1/n!$. Furthermore, the marginalization integral is overcounting the statistical weight of the expression by a factor of $(N-n)!$ which corresponds to all permutations of the outer particles. Therefore, multiplication of \eq{eq:MomentumFluxDivTermOneInnerOneOuterParticleII} {by ${{N}\choose{n}}$} yields a contribution 
%
\begin{equation}
\nabla_{\vec{p}_i}\cdot\Bigl(\Fmean(\vq_i) f_n(t, \VX^{n-1}, X_{i})\Bigr)
\end{equation}
to the evolution equation of $f_n$, i.e., the outer particles exert a meanfield force onto the particles within the open system. 


\subsection{Marginalizing w.r.t.\ outside particle position spaces}

Next we consider the physical space transport terms $\sum_{i=1}^N \nabla_{\vec{q}_i} \cdot \left({\red{\vec{v}_i}}F_N\right)$ {\green{from \eq{eq:liouvilleNPspecific}}} w.r.t.\ the outer particles. Consider first those terms in {the} sum for which $i \in 1, ..., n$. {For these}, 
\begin{equation}
{{N}\choose{n}} \int\limits_{(S^c)^{N-n}} 
\nabla_{\vec{q}_i} \cdot \left({\red{\vec{v}_i}} F_N(t, \VX^n, \VXi_n^N\right) \, d\VXi_n^N
=
\nabla_{\vec{q}_i} \cdot \left({\red{\vec{v}_i}} f_n\right)
\end{equation}
and we obtain the analogous transport term for the $n$-particle open system Liouville equation.
For ${\green{n+1}} \leq i \leq N$, however, one of the integrals will be over $\Xi_i \in S^c$, and for the
corresponding terms
\begin{equation}\label{eq:LEqnPosSpaceMarginal}
\begin{array}{rl} 
  & \dss {{N}\choose{n}} \sum\limits_{i={\green{n+1}}}^{N}
    \int\limits_{S^c} 
    \dss \int\limits_{(S^c)^{N-n-1}} 
    \nabla_{\vec{q}_i} \cdot \left({\red{\vec{v}_i}} F_N(t, \VX^n, \VXi_{n+1}^{N}, (\vq_i,\vp_i)\right)  
    \, d\VXi_{n+1}^N \ d \Xi_i
    \\[20pt]
  & = {{N}\choose{n}} (N-n) \ \dss \int\limits_{S^c} \nabla_{\vec{q}_i} \cdot 
      \biggl\{ \int\limits_{(S^c)^{N-n-1}} \left({\red{\vec{v}_i}} F_N(t, \VX^n, \VXi_{n+1}^{N}, (\vq_i,\vp_i)\right) \, d\VXi_{n+1}^{N}
      \biggr\}
      d \Xi_i
    \\
  & = - \dss (n+1) 
        \int\limits_{\partial \Omega} 
          \int\limits_{\reals^3}
            \left({\red{\vec{v}_i}} \cdot \vn\right) \ {\widehat f}_{n+1}(t, \VX^n, (\vq_i,\vp_i)\, 
          d^3 p_i\,
        d\sigma_i\,.
\end{array}
\end{equation}
Leaving the exact definition of ${\widehat f}_{n+1}$ to be discussed in the next paragraph, we notice that the factor $(N-n)$ in the second line arises because all the terms from the first line for $i \in {\green{n+1}}, ..., N$ are identical, and that the negative sign in the third line arises after we have applied Gau\ss' theorem to the integral over $\Omega_c$, whereas we have used the outer normal $\vn$ of $\Omega$ (not its complement) in the formula. Moreover, we have used that ${{N}\choose{n}} (N-n) = (n+1) {{N}\choose{n+1}}$ to let ${\widehat f}_{n+1}$ acquire the same scaling with $N$ and $n$ as ${f}_{n+1}$.

Next we need to distinguish the values of the marginalized density ${\widehat f}_{n+1}$ that are to be used for the two possible signs of $\left({\red{\vec{v}_i}} \cdot \vn\right)$. To this end, we {recall that the theory of characteristics states} that trajectories of the particle system carry with them their initial values of the density (see, e.g., Godlewski and Raviart (1996)\cite{GodlewskiRaviart1996}, section V, pp.~417–460). Therefore, when particles exit the open system, that is, $\left({\red{\vec{v}_i}} \cdot \vn\right) > 0$, then ${\widehat f}_{n+1} = f_{n+1}$, whereas for $\left({\red{\vec{v}_i}} \cdot \vn\right) < 0$ information is entering the open system from the outer universe, and {one} needs to decide upon a model to represent the statistical information transport in that case. {To emphasize this all-important degree of freedom in open system modelling, we follow two options in the sequel.}

A most straightforward approach assumes an equilibrium state of the universe that is statistically independent of the open system, and that amounts to letting ${\widehat f}_{n+1} = f_{n}f_1^{\circ}$, where $f_1^{\circ}$ is the single particle (equilibrium) density assumed for the reservoir. Thus, 
\begin{equation}\label{eq:OuterF}
{\widehat f}_{n+1} = 
\left\{
\begin{array}{l@{\quad}l}
f_{n+1} 
  & \left({\red{\vec{v}_i}} \cdot \vn > 0\right)
    \\
f_{n}f_1^{\circ}
  & \left({\red{\vec{v}_i}} \cdot \vn < 0\right) 
\end{array}
\right. \,.
\end{equation}
A second option is geared towards consistency with the thermodynamic and large system size limits for open systems. This amounts to imposing grand canonical distribution for state space trajectories that enter the open system from outside and reads
\begin{equation}\label{eq:OuterFGC}
{\widehat f}_{n+1} = 
\left\{
\begin{array}{l@{\quad}l}
f_{n+1} 
  & \left({\red{\vec{v}_i}} \cdot \vn > 0\right)
    \\
f_{n+1}^{\text{GC}}
  & \left({\red{\vec{v}_i}} \cdot \vn < 0\right) 
\end{array}
\right. \,.
\end{equation}
See an explicit representation of the grand canonical distribution $f_{n+1}^{\text{GC}}$ in \eq{probf1} below.


\subsection{Liouville-type equation for the $\VEC{n}$-state density}

From the previous subsections we collect the evolution equation
\begin{equation}\label{eq:LiouvilleTypeEquationI}
\frac{\partial f_n}{\partial t} 
+ \sum\limits_{i=1}^{n} \left(\nabla_{\vq_i}\cdot\left({\red{\vv_i}} f_n\right)  
+ \nabla_{\vp_i}  \cdot \left(\vF_i f_n\right)  \rule{0pt}{12pt}\right)
= \Psi_n + \Phi_{n}^{n+1}\,,
\end{equation}
where
\begin{equation}
\begin{array}{rcl}
\dss \vF_i 
  & =
     & \dss - \sum\limits_{j = 1; j\not=i}^{n}\nabla_{\vq_i} V(\vq_i - \vq_j)
\end{array}
\end{equation}
is the total force onto the $i$th particle exerted by the remaining $n-1$ particles
within the system through the potential interaction, and the coupling terms within
the $n$-particle hierarchy of distributions are
\begin{equation}\label{eq:LiouvilleHierarchy}
\begin{array}{rcl}
\Psi_n
  & =
    & \dss - \sum\limits_{i=1}^{n} \nabla_{\vec{p}_i}\cdot\Bigl(\Fmean(\vq_i) f_n(t, \VX^{i-1}, X_{i}, \VX_{i}^{n-i})\Bigr)
      \\
\Phi_{n}^{n+1}
  & =
    &  (n+1)\dss \int\limits_{\partial \Omega} 
          \int\limits_{({\red{\vec{v}_i}} \cdot \vn) > 0}
            \hspace{-5pt}
            \left({\red{\vec{v}_i}} \cdot \vn\right) \ 
            \left( {f}_{n+1}\left(t, \VX^{n},  (\vq_i,\vp_i)\right) 
                 - {f}_{n}\left(t, \VX^n\right){f}_{1}^{\circ}\left(\vq_i,-\vp_i\right) 
            \rule{0pt}{12pt}\right)\, 
          d^3 p_i\
        d\sigma_i
\end{array}
\end{equation}
where $f_1^{\circ}$ is the single particle (equilibrium) density assumed for the reservoir and
\begin{equation}
\Fmean(\vq_i) 
= 
- \int\limits_{S^c} 
  \nabla_{\vq_i}V(\vq_i- \vq_j) f^\circ_2(X_j | X_i) \,d X_j\,.
\end{equation}
%


\section{Grand Canonical Equilibrium Distribution function as verification of consistency of the model}
\label{gcverification}
In equilibrium, and under the hypothesis of short range interaction, that is $V(\vq_{i}-\vq_{j})$ decays fast as $|\vq_{i}-\vq_{j}|$ increases, the standard stationary Grand Canonical distribution is automatically obtained by the very definition of $f(t,\VX^{n})$ as verified next.\\
Let us consider the total system (i.e., the universe) with $N$ particles interacting through the Hamiltonian: $H_N 
= \sum_{i=1}^{N} \frac{\vp_i^2}{2M} 
+ \sum_{i=1}^{N} \sum_{j\not=i}^{N} \frac{1}{2} V(\vq_j-\vq_i)$, then its partition function reads:
\begin{equation}
  Q_{N}=\int_{U}\frac{1}{h^{3N}N!}e^{-\beta H_{N}}d^{N}\vq d^{N}\vp
  \label{partfun}
\end{equation}
with $h$ the Planck constant as usual, and $\beta=\frac{1}{k_{B}T}$, $k_{B}$ the Boltzmann constant.
The probability distribution for a given configuration $\VX^{N}$ is then written as:
\begin{equation}
  F_{N}(\VX^{N})=\frac{\frac{1}{h^{3N}N!}e^{-\beta H_{N}}}{Q_{N}}
  \label{distuniv}
\end{equation}
as a consequence, following the definition of \eq{eq:HierarchyDefinition}, in equilibrium one obtains the probability distribution of a subsystem of $n$ particles in a domain $\Omega$ as:
\begin{equation}
  f_n(t, \VX^n) =\frac{N!}{n!(N-n)!}\frac{\frac{1}{h^{3N}N!} e^{-\beta H_{n}}\int\limits_{(S^c)^{N-n}}e^{-\beta H_{N-n}+V_{cx}(\vq_{\Omega}-\vq_{\Omega^{c}})}\ d^{N-n}\Xi}{Q_{N}}
  \label{distom1}
\end{equation}
where $V_{cx}(\vq_{\Omega}-\vq_{\Omega^{c}})=\sum\limits_{i=1}^{n}\sum\limits_{j=n+1}^{N}V(\vq_{i}-\vq_{j})$ is the interaction potential between the $n$ particle inside and the $N-n$ particles outside.

{Under the hypotheses that (i) $N-n \gg n$, (ii) $ \Omega^{c} \gg \Omega$, and (iii) $V_{cx}(\vq_{\Omega}-\vq_{\Omega^{c}})$  is negligible compared to $\sum\limits_{i<j}^nV(\vq_i-\vq_j)$ for $\vq_i, \vq_j \in \Omega$ -- as justified, e.g., in case of short range potentials -- the total Hamiltonian can be factorized as: $H_{n}+H_{N-n}$. Then} the expression in \eq{distom1} can be reduced to:
\begin{equation}
  f_n(t, \VX^n)=\frac{1}{Q_{N}}\left[\frac{1}{n!(N-n)!}\frac{1}{h^{3n}h^{3(N-n)}}e^{-\beta H_{n}}\int\limits_{(S^c)^{N-n}}e^{-\beta H_{N-n}}\ d^{N-n}\Xi\right]
    \label{distom2}
\end{equation}
that is:
\begin{equation}
    f_n(t, \VX^n)=\frac{1}{h^{3n}n!}e^{-\beta H_{n}}\frac{Q_{N-n}}{Q_{N}}
      \label{distom3}
\end{equation}
with $Q_{N-n}=\frac{1}{h^{3(N-n)}(N-n)!}\int\limits_{(S^c)^{N-n}}e^{-\beta H_{N-n}}d^{N-n}\vq d^{N-n}\vp$.\\
    Next, following well established textbooks of statistical mechanics (see e.g. \cite{huangbook}), one has:
\begin{equation}
      \frac{Q_{N-n}}{Q_{N}}= \exp\left(-\beta[A(N-n,U\backslash\Omega, T)-A(N,U,T)]\right)
  \label{eqfreen}
\end{equation}
with $A$ the Helmholtz free energy at given number of particles, volume and temperature. {Since we assume $N\gg n$ and $\Omega_c \gg \Omega$, this yields: $A(N-n,U\backslash\Omega,T)-A(N,U,T)\approx n\mu+\Omega P$, with $\mu=\frac{\partial A}{\partial n}$ the chemical potential and $P=-\frac{\partial A}{\partial \Omega}$ the pressure.} As a consequence one gets the standard Grand Canonical probability distribution function:
\begin{equation}
  f_n(t, \VX^n)=  \frac{e^{-\beta (H_{n} - \mu n + P\Omega)}}{n!h^{3n}} 
  \label{probf1}
\end{equation}
and, as usual, the normalization condition:
\begin{equation}
\sum\limits_{n = 0}^{\infty} \
\int\limits_{\Omega^n} \int\limits_{(\reals^{3})^{n}} f_n(t,(\vq,\vp))\ d^n\vp\ \ d^n \vq = 1\,.
\end{equation}
implies the thermodynamic definition of the grand potential $\Phi=P\Omega=k_{B}T \ln Z(\mu,\Omega,T)$, with $Z(\mu,\Omega,T)=\sum_{n=0}^{\infty}\int\limits_{\Omega^n} \int\limits_{(\reals^{3})^{n}} \frac{1}{n!h^{3n}}e^{-\beta (H_{n} - \mu n)}\ d^n\vp\ \ d^n \vq$.

One can also verify that given the physical and mathematical hypothesis stated in the derivation above, the Grand Canonical probability distribution function of \eq{probf1}, satisfies \eq{eq:LiouvilleTypeEquationI}. In fact in equilibrium, one would have:
\begin{equation}
\sum\limits_{i=1}^{n} \left(\nabla_{\vq_i}\cdot\left({\red{\vv_i}} \frac{e^{-\beta (H_{n} - \mu n + P\Omega)}}{n!h^{3n}}\right) 
+ \nabla_{\vp_i}  \cdot \left(\vF_i  \frac{e^{-\beta (H_{n} - \mu n + P\Omega)}}{n!h^{3n}}\right)  \rule{0pt}{12pt}\right)
=\Psi_n + \Phi_{n}^{n+1}\,,
\label{rieqsub}
\end{equation}
{the left hand side equals zero, because $\nabla_{\vq_i}\cdot\left({\red{\vv_i}} \frac{e^{-\beta (H_{n} - \mu n + P\Omega)}}{n!h^{3n}}\right)=\vF_i\cdot{\red{\vv_{i}}}f_n(\VX^n)$ and $\nabla_{\vp_i}  \cdot \left(\vF_i  \frac{e^{-\beta (H_{n} - \mu n + P\Omega)}}{n!h^{3n}}\right)=-\vF_i\cdot{\red{\vv_{i}}}f_n(\VX^n)$. Next, the hypothesis that $V_{cx}(\vq_{\Omega}-\vq_{\Omega^{c}})$ is negligible, implies that $\Fmean(\vq_i)$ is negligible, thus in equilibrium, $\Psi_n$ is a negligible ``surface'' effect. Finally, for $ \Phi_{n}^{n+1}$, which describes the net flux of $n$-particle probability into/out of the open system, is trivially zero for grand canonical statistics of $f_{n+1}$ when we adopt the model from \eq{eq:OuterFGC}. 
For the other option in \eq{eq:OuterF} we see that exact cancellation of this term would only occur if $f^{\text{GC}}_{n+1} - f_n^{\text{GC}} f_1^\circ = 0$, which is generally not the case, since the Grand Canonical statistics does not factorize with respect to its position space dependence. Still, for large system size we can argue as for $\Psi_n$ and accept that the term is negligible to the extend that the surface-to-volume ratio of the open system is small.}

{%
The important aspect to be registered here is that in reality the ``true''  $ f_n(t, \VX^n)$ that one obtains in equilibrium, includes explicitly the surface effects -- as small as they might be -- without any hypothesis on the behavior of the potential. The Grand Canonical distribution, in turn, is only a physical approximation that becomes rigorous in the thermodynamic limit for large $n$ and large system size. In this sense it will be interesting to see, in section~\ref{sec:RelationBLvsOSM}, the differences between the present open system model and the BL model, which assumes $V_{cx}(\vq_{i}-\vq_{j})=0$ when $\vq_i \in \Omega$ and $\vq_j \in \Omega_c$ and relies on the impulsive interaction hypothesis. Similarly, in section~\ref{sec:OpenSystemVSAdResS} we discuss differences with computational algorithms in which $V_{cx}(\vq_{\Omega}-\vq_{\Omega^{c}})$ plays a relevant technical role.}


\section{Similarities and differences between the Bergmann-Lebowitz and the present open system model}
\label{sec:RelationBLvsOSM}
The key idea of the Bergmann-Lebowitz open system model \citep{leb1,leb2} is to 
discretize the interaction of an open system with its {environment. They} allow for
impulsive interactions of the system with statistically undisturbed reservoirs at discrete
points in time, the interactions being induced by a suitable interaction kernel. In essence, 
the interaction with the reservoir leads to a discrete transition of the system from a state 
characterized by $n$ particles $\VX^{n}$ to a state characterized by $m$ particles 
($\VY^m$). As a hypothesis/model, the macroscopic thermodynamic variables of the 
reservoir are not influenced by the system and molecules entering into the system {from 
the reservoir} can only have, in thermal equilibrium, velocities consistent with the temperature 
of the reservoir. 

These concepts are formalized via the the introduction of a stochastic
kernel, $K_{nm}(\VY^{m},\VX^{n})$, whose action allows for the transition from a microscopic state, $\VX^{n}$, to another state, $\VY^{m}$ (and vice versa). In essence, such kernel corresponds to the probability per unit time that the system at $\VX^{n}$ makes a transition to $\VY^{m}$, caused by the interaction between the system and the reservoir. The total system-reservoir interaction becomes: $\sum_{m=0}^{\infty}\int d\VY^m [K_{nm}(\VX^{n},\VY^{m})f_{m}(\VY^{m},t) - K_{mn}(\VY^{m},\VX^{n})f_n(\VX^{n},t)]$. {The time evolution of the probability, $f_n(\VX^{n},t)$, is described by a Liouville equation where the Poisson bracket on the r.h.s.\ is augmented by a term, in integral form, describing the exchange with the reservoir}:
\begin{multline}
\frac{\partial f_n(t,\VX^{n})}{\partial t}=\{f_n(\VX^{n},t),H_{n}(\VX^{n})\}+\\ 
+\sum_{m=0}^{\infty}\int d\VY^{m}[K_{nm}(\VX^{n},\VY^{m})f_{m}(\VY^{m},t)-K_{mn}(\VY^{m},\VX^{n})f_n(\VX^{n},t)].
\label{liouvext}
\end{multline}

The condition of flux balance defines the equilibrium for this model through
\begin{equation}
\sum_{m=0}^{\infty}\int d\VY^{m}[K_{nm}(\VX^{n},\VY^{m})f_{m}(\VY^{m},t)-K_{mn}(\VY^{m},\VX^{n})f_n(\VX^{n},t)]=0\,.
\label{fuba}
\end{equation}
and in equilibrium, the stationary solution for $f_n(\VX^{n})$ is the Grand Canonical probability density: $f_{n}(\VX^{n})=\frac{1}{Q}e^{-\beta H_{n}(\VX^{n})+ {\beta}\mu n}$ where $\beta=1/k_BT$ and $\mu$ the chemical potential. The condition of flux balance is necessary and sufficient for obtaining the Grand Canonical probability density as stationary solution. 

The key difference with the proposed model of this work is that we do not assume a discontinuous impulsive interactions of the system with the reservoir, but we allow for a dynamical continuous transition between different states {described by the Hamiltonian of the ``entire universe''.} Proper conditions of interaction {of the open system with the reservoir are derived analytically, and yield the terms, $\Psi_n + \Phi_{n}^{n+1}$, which link neighboring levels of the $n$-hierarchy of open system state space densities. These terms represent the flux of information in and out of the open system's domain, $\Omega$. An interesting application of the developed physico-mathematical open system model lies in its explanatory power of interpreting computational schemes that simulate an open subsystem with detailed atomistic particle representations embedded in a reduced complexity environment and the dynamical exchange of energy and particles between them.}  In the next section we will discuss in detail the connection between the proposed model and a well established computational algorithm for simulation of open molecular systems.


\section{Relation between the Adaptive Resolution Molecular Dynamics scheme and the proposed open system model}
\label{sec:OpenSystemVSAdResS}

Recent developments in computational physics have {contributed} to the construction of algorithms for molecular dynamics simulation of systems that exchange matter with a reservoir. Fig.~\ref{adapt} shows the set up of the latest version of the Adaptive Resolution Simulation method (AdResS) {together with} the {corresponding spatial partitioning} of the simulation box \cite{jcpabrupt,advtsres}. The {fully} atomistic region is interfaced with a transition region which in turn is interfaced with a region of non-interacting particles (tracers). Upon entering the transition region a tracer particle acquires an atomistic identity and while crossing this region the molecule equilibrates with the full atomistic environment. In the other direction, a molecule in the transition region entering into the tracer region becomes a non-interacting particle and loses its atomistic character; the process of changing identity occurs in an abrupt manner. 

The similarity between the theoretical model introduced in these notes and the computational implementation of AdResS can be brought forward noticing that the transition region of AdResS is essentially a region of the reservoir, where, according to the potential cutoff used, the terms $\Psi_n$ are defined and where the surface effects of $\Phi_{n}^{n+1}$ take place.
\begin{figure}
\includegraphics[height=5cm]{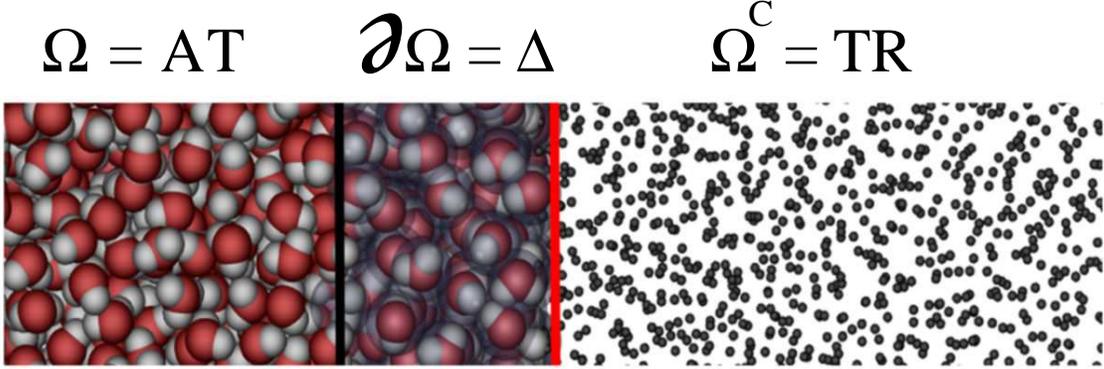}
  \caption{The partitioning of the simulation box in AdResS, an atomistic resolved region (AT), interfaced with a transition region ($\Delta$), which in turn is interfaced with a reservoir (TR) of non-interacting particles, called tracers. In analogy to the theoretical model proposed here, the AT region corresponds to the domain $\Omega$, the transition region, represents the ``surface'' $\partial\Omega$ and finally the reservoir TR corresponds to $\Omega^{c}$.}
\label{adapt}
\end{figure}
Moreover, in the transition region of AdResS an additional condition is imposed for the proper statistical exchange of particles between the system and reservoir. Such condition is derived from first principles of statistical mechanics with the aim of conserving the equilibrium statistical properties in AT region. {This condition is imposed in the AdResS model through an additional thermodynamic force that acts on particles only in the transition region \cite{prlgiov,prx}.}
 In conclusion, the AdResS scheme can be interpreted as a particle-based implementation of
the open system model proposed in section~\ref{sec:OpenSystemModel} above. Obviously, the 
atomistic region represents the open system, whereas the transition ($\Delta$) region is to 
represent the outside world. For distance-truncated interaction potentials, the only part of 
the outside world which the system sees is{, in fact,} a finite thickness layer of
particles that covers the cut-off length. Therefore, the thickness of the $\Delta$-region should 
be comparable to this cut-off length. Particles that reside beyond the $\Delta$-region, which
undergo simplified coarse-grained dynamics (they are just passive tracers in the present notes), 
merely serve as a particle reservoir that is needed to make sure the statistical balancing
mechanisms active in the $\Delta$-region always have a sufficient supply of particles.  
The main task of the $\Delta$-region in AdResS is to generate the desired outside world 
statistics. For a system as described above, whose dynamics involves only particle pair 
interactions, this means that the two-particle distribution $\fouter_2$ is to be established. 
Under the AdResS philosophy, this distribution is to emerge from explicit particle dynamics
in the $\Delta$-region which, due to the activity of a thermostat, can be expected to be
ergodic, and which, due to the action of the thermodynamic force, is guaranteed to at least 
provide the correct mean outside world particle density to particles within the open system.
Moreover, since in the $\Delta$-region the particles follow, except for these two effects, 
the same Hamiltonian dynamics as the particles inside, it is plausible that a particle
ensemble with the correct mean density and temperature will also adjust in space correctly
so as to reflect the radial distribution or particle pair position distributions. 
All the mechanisms of how the coupling between distributions of different particle numbers
interact and what are the proper boundary conditions carried by particles entering the 
domain are taken care of by AdResS automatically. This is because AdResS is particle based
and therefore implements the phase space transport of the distribution functions directly
in a ``Lagrangian fashion'' by generating stochastic trajectories.
{In summary, a clear comparison between the Bergmann-Lebowitz model and our model of open system regarding the formalization of real molecular simulation techniques, would lead to the following conclusions: the Bergmann-Lebowitz model is optimal for those simulation techniques based on stochastic sampling of the phase space configuration, (i.e. Monte Carlo methods), instead our model is optimal for techniques based on molecular dynamics schemes where equations of motions and molecular trajectories are determined at each instant of the simulation. In fact in the first case the system-reservoir coupling kernel, expressing the transition probability in the phase space, can be straightforwardly modeled with a Monte Carlo procedure, while in the second case the system-reservoir coupling term can be expressed through two-particle forces and particle-particle correlation functions, automatically calculated in a molecular dynamics scheme.}


\section{Conclusions}
\label{sec:Conclusions}

We have proposed a physico-mathematical model for open systems that exchange particles and energy with the external world. The procedure of reducing a large ``universe'' system to a smaller (sub)system is similar to the marginalization procedure of the BBGKY hierarchy scheme that defines a hierarchy of probability functions of the system. However, our model differs substantially from the BBGKY model because in the BBGKY model the marginalization is done independently of the position of the particles, {whereas} in our model the marginalization is applied with the {constraint} that particles are located outside a predefined region. As a consequence, in our case terms involving directly the partitioning of space (i.e. surface/boundary integrals) appear in the equations while this is not the case of the BBGKY model.

The equations obtained model a system of particles that continuously, in a dynamical fashion, exchanges energy and particles with the reservoir. This characteristic makes our model differ from {the well-established Bergmann-Lebowitz model of open systems, which is based instead on the hypothesis of discrete impulsive and stochastic} interaction between the system and the reservoir. The motivation for the development of our model lies on the fact that modern molecular dynamics schemes for open systems are based on a dynamical and continuous exchange of information with the exterior and thus they can only be approximately described by the Bergmann-Lebowitz model.

Instead, as we discuss in the paper, our model provides a closer formalization of the computational algorithms with the specific example of the mapping of our theoretical model to the AdResS method.

\acknowledgments{\green{This research has been funded by Deutsche Forschungsgemeinschaft (DFG) through grant CRC 1114 ``Scaling Cascade in Complex Systems,'' Project Number 235221301, Project C01 ``Adaptive coupling of scales in molecular dynamics and beyond to fluid dynamics.''}}


{\bf Data availability Statement}:
The data that supports the findings of this study are openly available on arXiv, reference number: 1907.07557
\bibliography{liouville}

\begin{thebibliography}{42}%
\makeatletter
\providecommand \@ifxundefined [1]{%
 \@ifx{#1\undefined}
}%
\providecommand \@ifnum [1]{%
 \ifnum #1\expandafter \@firstoftwo
 \else \expandafter \@secondoftwo
 \fi
}%
\providecommand \@ifx [1]{%
 \ifx #1\expandafter \@firstoftwo
 \else \expandafter \@secondoftwo
 \fi
}%
\providecommand \natexlab [1]{#1}%
\providecommand \enquote  [1]{``#1''}%
\providecommand \bibnamefont  [1]{#1}%
\providecommand \bibfnamefont [1]{#1}%
\providecommand \citenamefont [1]{#1}%
\providecommand \href@noop [0]{\@secondoftwo}%
\providecommand \href [0]{\begingroup \@sanitize@url \@href}%
\providecommand \@href[1]{\@@startlink{#1}\@@href}%
\providecommand \@@href[1]{\endgroup#1\@@endlink}%
\providecommand \@sanitize@url [0]{\catcode `\\12\catcode `\$12\catcode
  `\&12\catcode `\#12\catcode `\^12\catcode `\_12\catcode `\%12\relax}%
\providecommand \@@startlink[1]{}%
\providecommand \@@endlink[0]{}%
\providecommand \url  [0]{\begingroup\@sanitize@url \@url }%
\providecommand \@url [1]{\endgroup\@href {#1}{\urlprefix }}%
\providecommand \urlprefix  [0]{URL }%
\providecommand \Eprint [0]{\href }%
\providecommand \doibase [0]{http://dx.doi.org/}%
\providecommand \selectlanguage [0]{\@gobble}%
\providecommand \bibinfo  [0]{\@secondoftwo}%
\providecommand \bibfield  [0]{\@secondoftwo}%
\providecommand \translation [1]{[#1]}%
\providecommand \BibitemOpen [0]{}%
\providecommand \bibitemStop [0]{}%
\providecommand \bibitemNoStop [0]{.\EOS\space}%
\providecommand \EOS [0]{\spacefactor3000\relax}%
\providecommand \BibitemShut  [1]{\csname bibitem#1\endcsname}%
\let\auto@bib@innerbib\@empty
\bibitem [{\citenamefont {Quian}(2007)}]{quianrev}%
  \BibitemOpen
  \bibfield  {author} {\bibinfo {author} {\bibfnamefont {H.}~\bibnamefont
  {Quian}},\ }\bibfield  {title} {\enquote {\bibinfo {title} {Phosphorylation
  energy hypothesis: Open chemical systems and their biological functions},}\
  }\href@noop {} {\bibfield  {journal} {\bibinfo  {journal} {Annu. Rev. Phys.
  Chem.}\ }\textbf {\bibinfo {volume} {58}},\ \bibinfo {pages} {113} (\bibinfo
  {year} {2007})}\BibitemShut {NoStop}%
\bibitem [{\citenamefont {Bordin}\ \emph {et~al.}(2012)\citenamefont {Bordin},
  \citenamefont {Diehl}, \citenamefont {Barbosa},\ and\ \citenamefont
  {Levin}}]{bordin}%
  \BibitemOpen
  \bibfield  {author} {\bibinfo {author} {\bibfnamefont {J.}~\bibnamefont
  {Bordin}}, \bibinfo {author} {\bibfnamefont {A.}~\bibnamefont {Diehl}},
  \bibinfo {author} {\bibfnamefont {M.}~\bibnamefont {Barbosa}}, \ and\
  \bibinfo {author} {\bibfnamefont {Y.}~\bibnamefont {Levin}},\ }\bibfield
  {title} {\enquote {\bibinfo {title} {Ion fluxes through nanopores and
  transmembrane channels},}\ }\href@noop {} {\bibfield  {journal} {\bibinfo
  {journal} {Phys. Rev. E}\ }\textbf {\bibinfo {volume} {85}},\ \bibinfo
  {pages} {031914} (\bibinfo {year} {2012})}\BibitemShut {NoStop}%
\bibitem [{\citenamefont {Abramo}(2003)}]{abramo}%
  \BibitemOpen
  \bibfield  {author} {\bibinfo {author} {\bibfnamefont {A.}~\bibnamefont
  {Abramo}},\ }\bibfield  {title} {\enquote {\bibinfo {title} {Modeling
  electron transport in mosfet devices: Evolution and state of the art},}\
  }\href@noop {} {\bibfield  {journal} {\bibinfo  {journal} {in Advanced Device
  Modeling and Simulation, T.Grasser Ed. World Scientific}\ ,\ \bibinfo {pages}
  {1--27}} (\bibinfo {year} {2003})}\BibitemShut {NoStop}%
\bibitem [{\citenamefont {Politzer}(1996)}]{politzer}%
  \BibitemOpen
  \bibfield  {author} {\bibinfo {author} {\bibfnamefont {D.}~\bibnamefont
  {Politzer}},\ }\bibfield  {title} {\enquote {\bibinfo {title} {Condensate
  fluctuations of a trapped, ideal bose gas},}\ }\href@noop {} {\bibfield
  {journal} {\bibinfo  {journal} {Phys. Rev. A}\ }\textbf {\bibinfo {volume}
  {54}},\ \bibinfo {pages} {5048} (\bibinfo {year} {1996})}\BibitemShut
  {NoStop}%
\bibitem [{\citenamefont {Herzog}\ and\ \citenamefont
  {Olshanii}(1997)}]{herzog}%
  \BibitemOpen
  \bibfield  {author} {\bibinfo {author} {\bibfnamefont {C.}~\bibnamefont
  {Herzog}}\ and\ \bibinfo {author} {\bibfnamefont {M.}~\bibnamefont
  {Olshanii}},\ }\bibfield  {title} {\enquote {\bibinfo {title} {Trapped bose
  gas: The canonical versus grand canonical statistics},}\ }\href@noop {}
  {\bibfield  {journal} {\bibinfo  {journal} {Phys.Rev.A}\ }\textbf {\bibinfo
  {volume} {55}},\ \bibinfo {pages} {3254} (\bibinfo {year}
  {1997})}\BibitemShut {NoStop}%
\bibitem [{\citenamefont {Delle~Site}(2018{\natexlab{a}})}]{advtsel}%
  \BibitemOpen
  \bibfield  {author} {\bibinfo {author} {\bibfnamefont {L.}~\bibnamefont
  {Delle~Site}},\ }\bibfield  {title} {\enquote {\bibinfo {title} {Simulation
  of many-electron systems that exchange matter with the environmen},}\
  }\href@noop {} {\bibfield  {journal} {\bibinfo  {journal} {Adv.Th.Sim.}\
  }\textbf {\bibinfo {volume} {1}},\ \bibinfo {pages} {1800056} (\bibinfo
  {year} {2018}{\natexlab{a}})}\BibitemShut {NoStop}%
\bibitem [{\citenamefont {Frenkel}\ and\ \citenamefont
  {Smit}(2002)}]{Frenkelbook}%
  \BibitemOpen
  \bibfield  {author} {\bibinfo {author} {\bibfnamefont {D.}~\bibnamefont
  {Frenkel}}\ and\ \bibinfo {author} {\bibfnamefont {B.}~\bibnamefont {Smit}},\
  }\href@noop {} {\emph {\bibinfo {title} {Understanding Molecular
  Simulation}}},\ \bibinfo {edition} {2nd}\ ed.\ (\bibinfo  {publisher}
  {Academic Press},\ \bibinfo {address} {San Diego},\ \bibinfo {year}
  {2002})\BibitemShut {NoStop}%
\bibitem [{\citenamefont {Tuckerman}(2010)}]{tuckbook}%
  \BibitemOpen
  \bibfield  {author} {\bibinfo {author} {\bibfnamefont {M.~E.}\ \bibnamefont
  {Tuckerman}},\ }\href@noop {} {\emph {\bibinfo {title} {Statistical
  mechanics: theory and molecular simulation}}}\ (\bibinfo  {publisher} {Oxford
  University Press},\ \bibinfo {address} {New York},\ \bibinfo {year}
  {2010})\BibitemShut {NoStop}%
\bibitem [{\citenamefont {Praprotnik}, \citenamefont {Delle~Site},\ and\
  \citenamefont {Kremer}(2008)}]{adress2}%
  \BibitemOpen
  \bibfield  {author} {\bibinfo {author} {\bibfnamefont {M.}~\bibnamefont
  {Praprotnik}}, \bibinfo {author} {\bibfnamefont {L.}~\bibnamefont
  {Delle~Site}}, \ and\ \bibinfo {author} {\bibfnamefont {K.}~\bibnamefont
  {Kremer}},\ }\bibfield  {title} {\enquote {\bibinfo {title} {Multiscale
  simulation of soft matter: From scale bridging to adaptive resolution},}\
  }\href@noop {} {\bibfield  {journal} {\bibinfo  {journal} {Annu. Rev. Phys.
  Chem.}\ }\textbf {\bibinfo {volume} {59}},\ \bibinfo {pages} {545--571}
  (\bibinfo {year} {2008})}\BibitemShut {NoStop}%
\bibitem [{\citenamefont {Ensing}\ \emph {et~al.}(2007)\citenamefont {Ensing},
  \citenamefont {Nielsen}, \citenamefont {Moore}, \citenamefont {Klein},\ and\
  \citenamefont {Parrinello}}]{ensing}%
  \BibitemOpen
  \bibfield  {author} {\bibinfo {author} {\bibfnamefont {B.}~\bibnamefont
  {Ensing}}, \bibinfo {author} {\bibfnamefont {S.~O.}\ \bibnamefont {Nielsen}},
  \bibinfo {author} {\bibfnamefont {P.~B.}\ \bibnamefont {Moore}}, \bibinfo
  {author} {\bibfnamefont {M.~L.}\ \bibnamefont {Klein}}, \ and\ \bibinfo
  {author} {\bibfnamefont {M.}~\bibnamefont {Parrinello}},\ }\bibfield  {title}
  {\enquote {\bibinfo {title} {Energy conservation in adaptive hybrid
  atomistic/coarse-grain molecular dynamics},}\ }\href@noop {} {\bibfield
  {journal} {\bibinfo  {journal} {J. Chem. Theory Comput.}\ }\textbf {\bibinfo
  {volume} {3}},\ \bibinfo {pages} {1100} (\bibinfo {year} {2007})}\BibitemShut
  {NoStop}%
\bibitem [{\citenamefont {Heyden}\ and\ \citenamefont
  {Truhlar}(2008)}]{truhlar}%
  \BibitemOpen
  \bibfield  {author} {\bibinfo {author} {\bibfnamefont {A.}~\bibnamefont
  {Heyden}}\ and\ \bibinfo {author} {\bibfnamefont {D.~G.}\ \bibnamefont
  {Truhlar}},\ }\bibfield  {title} {\enquote {\bibinfo {title} {Conservative
  algorithm for an adaptive change of resolution in mixed
  atomistic/coarse-grained multiscale simulations},}\ }\href@noop {} {\bibfield
   {journal} {\bibinfo  {journal} {J. Chem. Theory Comput.}\ }\textbf {\bibinfo
  {volume} {4}},\ \bibinfo {pages} {217} (\bibinfo {year} {2008})}\BibitemShut
  {NoStop}%
\bibitem [{\citenamefont {Wagoner}\ and\ \citenamefont
  {Pande}(2013)}]{wagoner}%
  \BibitemOpen
  \bibfield  {author} {\bibinfo {author} {\bibfnamefont {J.}~\bibnamefont
  {Wagoner}}\ and\ \bibinfo {author} {\bibfnamefont {V.}~\bibnamefont
  {Pande}},\ }\bibfield  {title} {\enquote {\bibinfo {title} {Finite domain
  simulations with adaptive boundaries: Accurate potentials and nonequilibrium
  movesets},}\ }\href@noop {} {\bibfield  {journal} {\bibinfo  {journal}
  {J.Chem.Phys.}\ }\textbf {\bibinfo {volume} {139}},\ \bibinfo {pages}
  {234114} (\bibinfo {year} {2013})}\BibitemShut {NoStop}%
\bibitem [{\citenamefont {Mones}\ \emph {et~al.}(2015)\citenamefont {Mones},
  \citenamefont {Jones}, \citenamefont {G\"{o}tz}, \citenamefont {Laino},
  \citenamefont {Walker}, \citenamefont {Leimkuhler}, \citenamefont {Csany},\
  and\ \citenamefont {Bernstein}}]{csany}%
  \BibitemOpen
  \bibfield  {author} {\bibinfo {author} {\bibfnamefont {L.}~\bibnamefont
  {Mones}}, \bibinfo {author} {\bibfnamefont {A.}~\bibnamefont {Jones}},
  \bibinfo {author} {\bibfnamefont {A.}~\bibnamefont {G\"{o}tz}}, \bibinfo
  {author} {\bibfnamefont {T.}~\bibnamefont {Laino}}, \bibinfo {author}
  {\bibfnamefont {R.}~\bibnamefont {Walker}}, \bibinfo {author} {\bibfnamefont
  {B.}~\bibnamefont {Leimkuhler}}, \bibinfo {author} {\bibfnamefont
  {G.}~\bibnamefont {Csany}}, \ and\ \bibinfo {author} {\bibfnamefont
  {N.}~\bibnamefont {Bernstein}},\ }\bibfield  {title} {\enquote {\bibinfo
  {title} {The adaptive buffered force {QM}/{MM} method in the cp2k and amber
  software packages},}\ }\href@noop {} {\bibfield  {journal} {\bibinfo
  {journal} {J. Comp. Chem.}\ }\textbf {\bibinfo {volume} {36}},\ \bibinfo
  {pages} {633} (\bibinfo {year} {2015})}\BibitemShut {NoStop}%
\bibitem [{\citenamefont {Delle~Site}\ and\ \citenamefont
  {Praprotnik}(2017)}]{physrep}%
  \BibitemOpen
  \bibfield  {author} {\bibinfo {author} {\bibfnamefont {L.}~\bibnamefont
  {Delle~Site}}\ and\ \bibinfo {author} {\bibfnamefont {M.}~\bibnamefont
  {Praprotnik}},\ }\bibfield  {title} {\enquote {\bibinfo {title} {Molecular
  systems with open boundaries: Theory and simulation},}\ }\href {\doibase
  10.1016/j.physrep.2017.05.007} {\bibfield  {journal} {\bibinfo  {journal}
  {Phys. Rep.}\ }\textbf {\bibinfo {volume} {693}},\ \bibinfo {pages} {1--56}
  (\bibinfo {year} {2017})}\BibitemShut {NoStop}%
\bibitem [{\citenamefont {Delle~Site}(2018{\natexlab{b}})}]{cpcel}%
  \BibitemOpen
  \bibfield  {author} {\bibinfo {author} {\bibfnamefont {L.}~\bibnamefont
  {Delle~Site}},\ }\bibfield  {title} {\enquote {\bibinfo {title} {Grand
  canonical adaptive resolution simulation for molecules with electrons: A
  theoretical framework based on physical consistency},}\ }\href@noop {}
  {\bibfield  {journal} {\bibinfo  {journal} {Comp.Phys.Comm.}\ }\textbf
  {\bibinfo {volume} {222}},\ \bibinfo {pages} {94--101} (\bibinfo {year}
  {2018}{\natexlab{b}})}\BibitemShut {NoStop}%
\bibitem [{\citenamefont {Ciccotti}\ and\ \citenamefont
  {Delle~Site}(2019)}]{softmatt}%
  \BibitemOpen
  \bibfield  {author} {\bibinfo {author} {\bibfnamefont {G.}~\bibnamefont
  {Ciccotti}}\ and\ \bibinfo {author} {\bibfnamefont {L.}~\bibnamefont
  {Delle~Site}},\ }\bibfield  {title} {\enquote {\bibinfo {title} {The physics
  of open systems for the simulation of complex molecular environments in soft
  matter},}\ }\href@noop {} {\bibfield  {journal} {\bibinfo  {journal} {Soft
  Matter}\ }\textbf {\bibinfo {volume} {15}},\ \bibinfo {pages} {2114}
  (\bibinfo {year} {2019})}\BibitemShut {NoStop}%
\bibitem [{\citenamefont {Nielsen}, \citenamefont {Moore},\ and\ \citenamefont
  {Ensing}(2010)}]{ensingpol}%
  \BibitemOpen
  \bibfield  {author} {\bibinfo {author} {\bibfnamefont {S.~O.}\ \bibnamefont
  {Nielsen}}, \bibinfo {author} {\bibfnamefont {P.~B.}\ \bibnamefont {Moore}},
  \ and\ \bibinfo {author} {\bibfnamefont {B.}~\bibnamefont {Ensing}},\
  }\bibfield  {title} {\enquote {\bibinfo {title} {Adaptive multiscale
  molecular synamics of macromolecular fluids},}\ }\href@noop {} {\bibfield
  {journal} {\bibinfo  {journal} {Phys. Rev. Lett.}\ }\textbf {\bibinfo
  {volume} {105}},\ \bibinfo {pages} {237802} (\bibinfo {year}
  {2010})}\BibitemShut {NoStop}%
\bibitem [{\citenamefont {Delgado-Buscalioni}, \citenamefont {Sabli\'{c}},\
  and\ \citenamefont {Praprotnik}(2015)}]{mat1}%
  \BibitemOpen
  \bibfield  {author} {\bibinfo {author} {\bibfnamefont {R.}~\bibnamefont
  {Delgado-Buscalioni}}, \bibinfo {author} {\bibfnamefont {J.}~\bibnamefont
  {Sabli\'{c}}}, \ and\ \bibinfo {author} {\bibfnamefont {M.}~\bibnamefont
  {Praprotnik}},\ }\bibfield  {title} {\enquote {\bibinfo {title} {Open
  boundary molecular dynamics},}\ }\href@noop {} {\bibfield  {journal}
  {\bibinfo  {journal} {Eur. Phys. J. Special Topics}\ }\textbf {\bibinfo
  {volume} {224}},\ \bibinfo {pages} {2331--2349} (\bibinfo {year}
  {2015})}\BibitemShut {NoStop}%
\bibitem [{\citenamefont {Sabli\'{c}}, \citenamefont {Praprotnik},\ and\
  \citenamefont {Delgado-Buscalioni}(2016)}]{mat2}%
  \BibitemOpen
  \bibfield  {author} {\bibinfo {author} {\bibfnamefont {J.}~\bibnamefont
  {Sabli\'{c}}}, \bibinfo {author} {\bibfnamefont {M.}~\bibnamefont
  {Praprotnik}}, \ and\ \bibinfo {author} {\bibfnamefont {R.}~\bibnamefont
  {Delgado-Buscalioni}},\ }\bibfield  {title} {\enquote {\bibinfo {title} {Open
  boundary molecular dynamics of sheared star-polymer melts},}\ }\href@noop {}
  {\bibfield  {journal} {\bibinfo  {journal} {Soft Matter}\ }\textbf {\bibinfo
  {volume} {12}},\ \bibinfo {pages} {2416--2439} (\bibinfo {year}
  {2016})}\BibitemShut {NoStop}%
\bibitem [{\citenamefont {Fiorentini}\ \emph {et~al.}(2017)\citenamefont
  {Fiorentini}, \citenamefont {Kremer}, \citenamefont {Potestio},\ and\
  \citenamefont {Fogarty}}]{krem1}%
  \BibitemOpen
  \bibfield  {author} {\bibinfo {author} {\bibfnamefont {R.}~\bibnamefont
  {Fiorentini}}, \bibinfo {author} {\bibfnamefont {K.}~\bibnamefont {Kremer}},
  \bibinfo {author} {\bibfnamefont {R.}~\bibnamefont {Potestio}}, \ and\
  \bibinfo {author} {\bibfnamefont {A.~C.}\ \bibnamefont {Fogarty}},\
  }\bibfield  {title} {\enquote {\bibinfo {title} {Using force-based adaptive
  resolution simulations to calculate solvation free energies of amino acid
  sidechain analogues},}\ }\href@noop {} {\bibfield  {journal} {\bibinfo
  {journal} {J. Chem. Phys.}\ }\textbf {\bibinfo {volume} {146}},\ \bibinfo
  {pages} {244113} (\bibinfo {year} {2017})}\BibitemShut {NoStop}%
\bibitem [{\citenamefont {Krekeler}\ and\ \citenamefont
  {Delle~Site}(2017)}]{krek1}%
  \BibitemOpen
  \bibfield  {author} {\bibinfo {author} {\bibfnamefont {C.}~\bibnamefont
  {Krekeler}}\ and\ \bibinfo {author} {\bibfnamefont {L.}~\bibnamefont
  {Delle~Site}},\ }\bibfield  {title} {\enquote {\bibinfo {title} {Towards open
  boundary molecular dynamics simulation of ionic liquids},}\ }\href@noop {}
  {\bibfield  {journal} {\bibinfo  {journal} {Phys. Chem. Chem. Phys.}\
  }\textbf {\bibinfo {volume} {19}},\ \bibinfo {pages} {4701} (\bibinfo {year}
  {2017})}\BibitemShut {NoStop}%
\bibitem [{\citenamefont {Netz}, \citenamefont {Potestio},\ and\ \citenamefont
  {Kremer}(2016)}]{krem2}%
  \BibitemOpen
  \bibfield  {author} {\bibinfo {author} {\bibfnamefont {P.~A.}\ \bibnamefont
  {Netz}}, \bibinfo {author} {\bibfnamefont {R.}~\bibnamefont {Potestio}}, \
  and\ \bibinfo {author} {\bibfnamefont {K.}~\bibnamefont {Kremer}},\
  }\bibfield  {title} {\enquote {\bibinfo {title} {Adaptive resolution
  simulation of oligonucleotides},}\ }\href@noop {} {\bibfield  {journal}
  {\bibinfo  {journal} {J. Chem. Phys.}\ }\textbf {\bibinfo {volume} {145}},\
  \bibinfo {pages} {234101} (\bibinfo {year} {2016})}\BibitemShut {NoStop}%
\bibitem [{\citenamefont {Shadrack~Jabes}\ \emph {et~al.}(2018)\citenamefont
  {Shadrack~Jabes}, \citenamefont {Krekeler}, \citenamefont {Klein},\ and\
  \citenamefont {Delle~Site}}]{shad1}%
  \BibitemOpen
  \bibfield  {author} {\bibinfo {author} {\bibfnamefont {B.}~\bibnamefont
  {Shadrack~Jabes}}, \bibinfo {author} {\bibfnamefont {C.}~\bibnamefont
  {Krekeler}}, \bibinfo {author} {\bibfnamefont {R.}~\bibnamefont {Klein}}, \
  and\ \bibinfo {author} {\bibfnamefont {L.}~\bibnamefont {Delle~Site}},\
  }\bibfield  {title} {\enquote {\bibinfo {title} {Probing spatial locality in
  ionic liquids with the grand canonical adaptive resolution molecular dynamics
  technique},}\ }\href@noop {} {\bibfield  {journal} {\bibinfo  {journal} {J.
  Chem. Phys.}\ }\textbf {\bibinfo {volume} {148}},\ \bibinfo {pages} {193804}
  (\bibinfo {year} {2018})}\BibitemShut {NoStop}%
\bibitem [{\citenamefont {Shadrack~Jabes}, \citenamefont {Klein},\ and\
  \citenamefont {Delle~Site}(2018)}]{shad2}%
  \BibitemOpen
  \bibfield  {author} {\bibinfo {author} {\bibfnamefont {B.}~\bibnamefont
  {Shadrack~Jabes}}, \bibinfo {author} {\bibfnamefont {R.}~\bibnamefont
  {Klein}}, \ and\ \bibinfo {author} {\bibfnamefont {L.}~\bibnamefont
  {Delle~Site}},\ }\bibfield  {title} {\enquote {\bibinfo {title} {Structural
  locality and early stage of aggregation of micelles in water: An adaptive
  resolution molecular dynamics study},}\ }\href@noop {} {\bibfield  {journal}
  {\bibinfo  {journal} {Adv. Theor. Simul.}\ }\textbf {\bibinfo {volume} {1}},\
  \bibinfo {pages} {1800025} (\bibinfo {year} {2018})}\BibitemShut {NoStop}%
\bibitem [{\citenamefont {Shadrack~Jabes}\ and\ \citenamefont
  {Delle~Site}(2018)}]{shad3}%
  \BibitemOpen
  \bibfield  {author} {\bibinfo {author} {\bibfnamefont {B.}~\bibnamefont
  {Shadrack~Jabes}}\ and\ \bibinfo {author} {\bibfnamefont {L.}~\bibnamefont
  {Delle~Site}},\ }\bibfield  {title} {\enquote {\bibinfo {title} {Nanoscale
  domains in ionic liquids: A statistical mechanics definition for molecular
  dynamics studies},}\ }\href@noop {} {\bibfield  {journal} {\bibinfo
  {journal} {J. Chem. Phys.}\ }\textbf {\bibinfo {volume} {149}},\ \bibinfo
  {pages} {184502} (\bibinfo {year} {2018})}\BibitemShut {NoStop}%
\bibitem [{\citenamefont {Zavadlav}\ \emph {et~al.}(2018)\citenamefont
  {Zavadlav}, \citenamefont {Sablic}, \citenamefont {Podgornik},\ and\
  \citenamefont {Praprotnik}}]{mat3}%
  \BibitemOpen
  \bibfield  {author} {\bibinfo {author} {\bibfnamefont {J.}~\bibnamefont
  {Zavadlav}}, \bibinfo {author} {\bibfnamefont {J.}~\bibnamefont {Sablic}},
  \bibinfo {author} {\bibfnamefont {R.}~\bibnamefont {Podgornik}}, \ and\
  \bibinfo {author} {\bibfnamefont {M.}~\bibnamefont {Praprotnik}},\ }\bibfield
   {title} {\enquote {\bibinfo {title} {Open-boundary molecular dynamics of a
  dna molecule in a hybrid explicit/implicit salt solution},}\ }\href@noop {}
  {\bibfield  {journal} {\bibinfo  {journal} {Biophys.J.}\ }\textbf {\bibinfo
  {volume} {114}},\ \bibinfo {pages} {2352} (\bibinfo {year}
  {2018})}\BibitemShut {NoStop}%
\bibitem [{\citenamefont {Wagoner}\ and\ \citenamefont
  {Pande}(2018)}]{pandemonium}%
  \BibitemOpen
  \bibfield  {author} {\bibinfo {author} {\bibfnamefont {J.}~\bibnamefont
  {Wagoner}}\ and\ \bibinfo {author} {\bibfnamefont {V.}~\bibnamefont
  {Pande}},\ }\bibfield  {title} {\enquote {\bibinfo {title} {Adaptive
  boundaries in multiscale simulations},}\ }\href@noop {} {\bibfield  {journal}
  {\bibinfo  {journal} {J.Chem.Phys.}\ }\textbf {\bibinfo {volume} {148}},\
  \bibinfo {pages} {141104} (\bibinfo {year} {2018})}\BibitemShut {NoStop}%
\bibitem [{\citenamefont {Agarwal}\ \emph {et~al.}(2015)\citenamefont
  {Agarwal}, \citenamefont {Zhu}, \citenamefont {Hartmann}, \citenamefont
  {Wang},\ and\ \citenamefont {Delle~Site}}]{njp}%
  \BibitemOpen
  \bibfield  {author} {\bibinfo {author} {\bibfnamefont {A.}~\bibnamefont
  {Agarwal}}, \bibinfo {author} {\bibfnamefont {J.}~\bibnamefont {Zhu}},
  \bibinfo {author} {\bibfnamefont {C.}~\bibnamefont {Hartmann}}, \bibinfo
  {author} {\bibfnamefont {H.}~\bibnamefont {Wang}}, \ and\ \bibinfo {author}
  {\bibfnamefont {L.}~\bibnamefont {Delle~Site}},\ }\bibfield  {title}
  {\enquote {\bibinfo {title} {Molecular dynamics in a grand ensemble:
  {B}ergmann--{L}ebowitz model and adaptive resolution simulation},}\
  }\href@noop {} {\bibfield  {journal} {\bibinfo  {journal} {New. J. Phys.}\
  }\textbf {\bibinfo {volume} {17}},\ \bibinfo {pages} {083042} (\bibinfo
  {year} {2015})}\BibitemShut {NoStop}%
\bibitem [{\citenamefont {Agarwal}\ and\ \citenamefont
  {Delle~Site}(2015)}]{jcppi}%
  \BibitemOpen
  \bibfield  {author} {\bibinfo {author} {\bibfnamefont {A.}~\bibnamefont
  {Agarwal}}\ and\ \bibinfo {author} {\bibfnamefont {L.}~\bibnamefont
  {Delle~Site}},\ }\bibfield  {title} {\enquote {\bibinfo {title} {Path
  integral molecular dynamics within the grand canonical-like adaptive
  resolution technique: Simulation of liquid water},}\ }\href@noop {}
  {\bibfield  {journal} {\bibinfo  {journal} {J. Chem. Phys.}\ }\textbf
  {\bibinfo {volume} {143}},\ \bibinfo {pages} {094102} (\bibinfo {year}
  {2015})}\BibitemShut {NoStop}%
\bibitem [{\citenamefont {Delle~Site}(2016)}]{preliou}%
  \BibitemOpen
  \bibfield  {author} {\bibinfo {author} {\bibfnamefont {L.}~\bibnamefont
  {Delle~Site}},\ }\bibfield  {title} {\enquote {\bibinfo {title} {Formulation
  of liouville's theorem for grand ensemble molecular simulations},}\
  }\href@noop {} {\bibfield  {journal} {\bibinfo  {journal} {Phys.Rev.E}\
  }\textbf {\bibinfo {volume} {93}},\ \bibinfo {pages} {022130} (\bibinfo
  {year} {2016})}\BibitemShut {NoStop}%
\bibitem [{\citenamefont {Agarwal}\ and\ \citenamefont
  {Delle~Site}(2016)}]{cpcpi}%
  \BibitemOpen
  \bibfield  {author} {\bibinfo {author} {\bibfnamefont {A.}~\bibnamefont
  {Agarwal}}\ and\ \bibinfo {author} {\bibfnamefont {L.}~\bibnamefont
  {Delle~Site}},\ }\bibfield  {title} {\enquote {\bibinfo {title}
  {Grand-canonical adaptive resolution centroid molecular dynamics:
  Implementation and application},}\ }\href@noop {} {\bibfield  {journal}
  {\bibinfo  {journal} {Comp. Phys. Comm.}\ }\textbf {\bibinfo {volume}
  {206}},\ \bibinfo {pages} {26} (\bibinfo {year} {2016})}\BibitemShut
  {NoStop}%
\bibitem [{\citenamefont {Agarwal}, \citenamefont {Clementi},\ and\
  \citenamefont {Delle~Site}(2017)}]{pccppi}%
  \BibitemOpen
  \bibfield  {author} {\bibinfo {author} {\bibfnamefont {A.}~\bibnamefont
  {Agarwal}}, \bibinfo {author} {\bibfnamefont {C.}~\bibnamefont {Clementi}}, \
  and\ \bibinfo {author} {\bibfnamefont {L.}~\bibnamefont {Delle~Site}},\
  }\bibfield  {title} {\enquote {\bibinfo {title} {Path integral-{GC}-{AdResS}
  simulation of a large hydrophobic solute in water: A tool to investigate the
  interplay between local microscopic structures and quantum delocalization of
  atoms in space},}\ }\href@noop {} {\bibfield  {journal} {\bibinfo  {journal}
  {Phys. Chem. Chem. Phys.}\ }\textbf {\bibinfo {volume} {19}},\ \bibinfo
  {pages} {13030} (\bibinfo {year} {2017})}\BibitemShut {NoStop}%
\bibitem [{\citenamefont {Praprotnik}, \citenamefont {Delle~Site},\ and\
  \citenamefont {Kremer}(2005)}]{adress1}%
  \BibitemOpen
  \bibfield  {author} {\bibinfo {author} {\bibfnamefont {M.}~\bibnamefont
  {Praprotnik}}, \bibinfo {author} {\bibfnamefont {L.}~\bibnamefont
  {Delle~Site}}, \ and\ \bibinfo {author} {\bibfnamefont {K.}~\bibnamefont
  {Kremer}},\ }\bibfield  {title} {\enquote {\bibinfo {title} {Adaptive
  resolution molecular-dynamics simulation: Changing the degrees of freedom on
  the fly},}\ }\href@noop {} {\bibfield  {journal} {\bibinfo  {journal} {J.
  Chem. Phys.}\ }\textbf {\bibinfo {volume} {123}},\ \bibinfo {pages} {224106}
  (\bibinfo {year} {2005})}\BibitemShut {NoStop}%
\bibitem [{\citenamefont {Lebowitz}\ and\ \citenamefont
  {Bergmann}(1957)}]{leb1}%
  \BibitemOpen
  \bibfield  {author} {\bibinfo {author} {\bibfnamefont {J.}~\bibnamefont
  {Lebowitz}}\ and\ \bibinfo {author} {\bibfnamefont {P.}~\bibnamefont
  {Bergmann}},\ }\bibfield  {title} {\enquote {\bibinfo {title} {Irreversible
  {G}ibbsian ensembles},}\ }\href@noop {} {\bibfield  {journal} {\bibinfo
  {journal} {Ann. Phys.}\ }\textbf {\bibinfo {volume} {1}},\ \bibinfo {pages}
  {1} (\bibinfo {year} {1957})}\BibitemShut {NoStop}%
\bibitem [{\citenamefont {Bergmann}\ and\ \citenamefont
  {Lebowitz}(1955)}]{leb2}%
  \BibitemOpen
  \bibfield  {author} {\bibinfo {author} {\bibfnamefont {P.}~\bibnamefont
  {Bergmann}}\ and\ \bibinfo {author} {\bibfnamefont {J.}~\bibnamefont
  {Lebowitz}},\ }\bibfield  {title} {\enquote {\bibinfo {title} {New approach
  to nonequilibrium processes},}\ }\href@noop {} {\bibfield  {journal}
  {\bibinfo  {journal} {Phys. Rev.}\ }\textbf {\bibinfo {volume} {99}},\
  \bibinfo {pages} {578} (\bibinfo {year} {1955})}\BibitemShut {NoStop}%
\bibitem [{\citenamefont {Bogoliubov}(1946)}]{bog3}%
  \BibitemOpen
  \bibfield  {author} {\bibinfo {author} {\bibfnamefont {N.}~\bibnamefont
  {Bogoliubov}},\ }\bibfield  {title} {\enquote {\bibinfo {title} {Kinetic
  equations},}\ }\href@noop {} {\bibfield  {journal} {\bibinfo  {journal}
  {Journal of Physics USSR}\ }\textbf {\bibinfo {volume} {10}},\ \bibinfo
  {pages} {265} (\bibinfo {year} {1946})}\BibitemShut {NoStop}%
\bibitem [{\citenamefont {Huang}(1986)}]{huangbook}%
  \BibitemOpen
  \bibfield  {author} {\bibinfo {author} {\bibfnamefont {K.}~\bibnamefont
  {Huang}},\ }\href@noop {} {\emph {\bibinfo {title} {Statistical mechanics}}}\
  (\bibinfo  {publisher} {Wiley},\ \bibinfo {year} {1986})\BibitemShut
  {NoStop}%
\bibitem [{\citenamefont {Godlewski}\ and\ \citenamefont
  {Raviart}(1996)}]{GodlewskiRaviart1996}%
  \BibitemOpen
  \bibfield  {author} {\bibinfo {author} {\bibfnamefont {E.}~\bibnamefont
  {Godlewski}}\ and\ \bibinfo {author} {\bibfnamefont {P.}~\bibnamefont
  {Raviart}},\ }\href@noop {} {\emph {\bibinfo {title} {Numerical approximation
  of the hyperbolic systems of conservation laws}}},\ Applied mathematical
  Sciences\ (\bibinfo  {publisher} {Springer-Verlag},\ \bibinfo {address} {New
  York},\ \bibinfo {year} {1996})\BibitemShut {NoStop}%
\bibitem [{\citenamefont {Krekeler}\ \emph {et~al.}(2018)\citenamefont
  {Krekeler}, \citenamefont {Agarwal}, \citenamefont {Junghans}, \citenamefont
  {Praprotnik},\ and\ \citenamefont {Delle~Site}}]{jcpabrupt}%
  \BibitemOpen
  \bibfield  {author} {\bibinfo {author} {\bibfnamefont {C.}~\bibnamefont
  {Krekeler}}, \bibinfo {author} {\bibfnamefont {A.}~\bibnamefont {Agarwal}},
  \bibinfo {author} {\bibfnamefont {C.}~\bibnamefont {Junghans}}, \bibinfo
  {author} {\bibfnamefont {M.}~\bibnamefont {Praprotnik}}, \ and\ \bibinfo
  {author} {\bibfnamefont {L.}~\bibnamefont {Delle~Site}},\ }\bibfield  {title}
  {\enquote {\bibinfo {title} {Adaptive resolution molecular dynamics
  technique: Down to the essential},}\ }\href@noop {} {\bibfield  {journal}
  {\bibinfo  {journal} {J.Chem.Phys.}\ }\textbf {\bibinfo {volume} {149}},\
  \bibinfo {pages} {024104} (\bibinfo {year} {2018})}\BibitemShut {NoStop}%
\bibitem [{\citenamefont {Delle~Site}\ \emph {et~al.}(2019)\citenamefont
  {Delle~Site}, \citenamefont {Krekeler}, \citenamefont {Whittaker},
  \citenamefont {Agarwal}, \citenamefont {Klein},\ and\ \citenamefont
  {H\"{o}fling}}]{advtsres}%
  \BibitemOpen
  \bibfield  {author} {\bibinfo {author} {\bibfnamefont {L.}~\bibnamefont
  {Delle~Site}}, \bibinfo {author} {\bibfnamefont {C.}~\bibnamefont
  {Krekeler}}, \bibinfo {author} {\bibfnamefont {J.}~\bibnamefont {Whittaker}},
  \bibinfo {author} {\bibfnamefont {A.}~\bibnamefont {Agarwal}}, \bibinfo
  {author} {\bibfnamefont {R.}~\bibnamefont {Klein}}, \ and\ \bibinfo {author}
  {\bibfnamefont {F.}~\bibnamefont {H\"{o}fling}},\ }\bibfield  {title}
  {\enquote {\bibinfo {title} {Molecular dynamics of open systems: construction
  of a mean-field particle reservoir},}\ }\href@noop {} {\bibfield  {journal}
  {\bibinfo  {journal} {Adv.Th.Sim.}\ }\textbf {\bibinfo {volume} {2}},\
  \bibinfo {pages} {1900014} (\bibinfo {year} {2019})}\BibitemShut {NoStop}%
\bibitem [{\citenamefont {Fritsch}\ \emph {et~al.}(2012)\citenamefont
  {Fritsch}, \citenamefont {Poblete}, \citenamefont {Junghans}, \citenamefont
  {Ciccotti}, \citenamefont {Delle~Site}, ,\ and\ \citenamefont
  {Kremer}}]{prlgiov}%
  \BibitemOpen
  \bibfield  {author} {\bibinfo {author} {\bibfnamefont {S.}~\bibnamefont
  {Fritsch}}, \bibinfo {author} {\bibfnamefont {S.}~\bibnamefont {Poblete}},
  \bibinfo {author} {\bibfnamefont {C.}~\bibnamefont {Junghans}}, \bibinfo
  {author} {\bibfnamefont {G.}~\bibnamefont {Ciccotti}}, \bibinfo {author}
  {\bibfnamefont {L.}~\bibnamefont {Delle~Site}}, , \ and\ \bibinfo {author}
  {\bibfnamefont {K.}~\bibnamefont {Kremer}},\ }\bibfield  {title} {\enquote
  {\bibinfo {title} {Adaptive resolution molecular dynamics simulation through
  coupling to an internal particle reservoir},}\ }\href@noop {} {\bibfield
  {journal} {\bibinfo  {journal} {Phys. Rev. Lett.}\ }\textbf {\bibinfo
  {volume} {108}},\ \bibinfo {pages} {170602} (\bibinfo {year}
  {2012})}\BibitemShut {NoStop}%
\bibitem [{\citenamefont {Wang}\ \emph {et~al.}(2013)\citenamefont {Wang},
  \citenamefont {Hartmann}, \citenamefont {Sch{\"u}tte}, ,\ and\ \citenamefont
  {Delle~Site}}]{prx}%
  \BibitemOpen
  \bibfield  {author} {\bibinfo {author} {\bibfnamefont {H.}~\bibnamefont
  {Wang}}, \bibinfo {author} {\bibfnamefont {C.}~\bibnamefont {Hartmann}},
  \bibinfo {author} {\bibfnamefont {C.}~\bibnamefont {Sch{\"u}tte}}, , \ and\
  \bibinfo {author} {\bibfnamefont {L.}~\bibnamefont {Delle~Site}},\ }\bibfield
   {title} {\enquote {\bibinfo {title} {Grand-canonical-like molecular-dynamics
  simulations by using an adaptive-resolution technique},}\ }\href@noop {}
  {\bibfield  {journal} {\bibinfo  {journal} {Phys. Rev. X}\ }\textbf {\bibinfo
  {volume} {3}},\ \bibinfo {pages} {011018} (\bibinfo {year}
  {2013})}\BibitemShut {NoStop}%
\end{thebibliography}%
\end{document}